\documentclass[12pt]{article}
   \usepackage{amssymb}
   \usepackage{amsmath}
   \usepackage{subfigure}
         \usepackage{epsfig}
         \def\href#1#2{#2}
         \def\IP{\relax{\rm I\kern-.18em P}}
         \newcommand{\beq}{\begin{equation}} 
         \newcommand{\eeq}{\end{equation}}
         \newcommand{\beqa}{\begin{eqnarray}}
         \newcommand{\eeqa}{\end{eqnarray}}

         \def\be{ \begin{equation}}\def\ee{ \end{equation}}
         \def\ba{ \begin{eqnarray}}\def\ea{ \end{eqnarray}}

         \def\cedille#1{\setbox0=\hbox{#1}\ifdim\ht0=1ex \accent'30 #1%
          \else{\ooalign{\hidewidth\char'30\hidewidth\crcr\unbox0}}\fi}
         
\begin{document}

{}~ \hfill\vbox{\hbox{hep-th/0310138}
\hbox{UUITP-16/03}
}\break

\vskip 1.cm

\centerline{\large \bf On Effective Actions for the Bosonic Tachyon}
\vspace*{1.5ex}

\vspace*{4.0ex}

\centerline{\large \rm M.\ Smedb\"ack\footnote{
mikael.smedback@teorfys.uu.se}}
\vspace*{2.5ex}
\centerline{\large \it Department of Theoretical Physics}
\centerline{\large \it Box 803, SE-751 08 Uppsala, Sweden}
\vspace*{3.0ex}

\centerline{\large November 9, 2003}
\vspace*{3.0ex}

\vspace*{4.5ex}
\medskip
\bigskip\bigskip
\centerline {\bf Abstract}

         We extend the analysis of hep-th/0304045 to the
         bosonic case and find the one-derivative
         effective action valid in the vicinity of
         rolling tachyons with an energy not larger than
         that of the original D-brane. 
         For on-shell tachyons rolling down the
         well-behaved side of the potential in this theory, the energy
         is conserved and the pressure eventually decreases 
         exponentially. For tachyons rolling down the
         ``wrong'' side, the pressure instead blows up
         in a finite time.
         
\bigskip

\vfill \eject
\baselineskip=17pt


         \tableofcontents

         \section{Introduction}
         \setcounter{equation}{0}

         From an analysis
         \cite{SenRolling, SenMatter}
         using boundary CFT (Conformal Field Theory) techniques
         \cite{CKLM, PT, Schomerus1, SenDescent}
         it has been possible to deduce different
         properties of the open-string tachyon\footnote{
         Recent progress on both open-string and closed-string tachyon dynamics include 
         \cite{ZZ, Str1, Str2, Str3, Str4, KMS, Kluson2, Schomerus2, Kluson3, Kluson4, Schomerus3, Kluson5}.}. 
         It was subsequently attempted to reproduce these properties
         from an effective field theory. The 
         ``standard form'' Lagrangian\footnote{In this paper, the flat 
         metric is
         $\eta_{\mu \nu} = (-1,+1,\cdots,+1)$ and $\alpha'=1$.}
         \begin{equation}\label{sec1_standard}
           L = -\frac{1}{\cosh \left( \frac{\alpha \tilde{T}}{2} \right)}
               \sqrt{1 + \partial_\mu \tilde{T} \partial^\mu \tilde{T}}
         \end{equation}
         was proposed \cite{SenRolling, SenMatter, SenTT, SenTimeEvol, 
         BergshoeffAction, GarousiAction, KlusonAction}, 
         based on its ability to reproduce some of
         the properties of the tachyon known from the boundary CFT
         analysis. Here, $\alpha=1$ describes the bosonic tachyon,
         while $\alpha=\sqrt{2}$ corresponds to the superstring
         case.

         Recently, Kutasov and Niarchos \cite{KutNi} 
         managed to
         derive this action in the superstring case with an ansatz
         of the type
         $L(T,\partial_\mu T \partial^\mu T)$ using only
         two requirements:
         First, that the action should allow the known tachyon solutions.
         And second, that it should be consistent with certain results
         from boundary SFT (String Field Theory)\footnote{
         A similar argument appeared already in \cite{LLM}.}. 

         The purpose of this paper is to try to extend the
         derivation to the bosonic case. We give the
         details of this derivation in section \ref{sec2}.
         In section \ref{sec3} we investigate what the
         properties of the resulting action are, and
         finally in section \ref{sec4} the validity
         region of the action is discussed.

         \section{Deriving an Effective Action for the Tachyon}\label{sec2} 
         \setcounter{equation}{0} 

         In this section, we will derive the effective action.
         In section \ref{sec2_conditions} we will explain the conditions
         which are used to determine the Lagrangian. In section
         \ref{sec2_derivation} we will then use these conditions to
         derive the action. The derivation is analogous to that of
         \cite{KutNi} for the superstring case.

         \subsection{Conditions}\label{sec2_conditions}
         Let us assume that we have a Lagrangian of the form 
         $L=L(T,\partial_{\mu}T \partial^{\mu}T)$. 
         
         We begin by discussing the EOM condition arising from
         requiring that the equations of motion be
         satisfied by the rolling tachyon solution, using the
         formalism of \cite{LS}.
         We write
         \begin{equation}\label{sec2_L}
           L = L_{even} + L_{odd}
         \end{equation}
         where the even part is
         \begin{equation}\label{sec2_Leven}
           L_{even} = \sum_{n=0}^{n=\infty} L_{2n}
         \end{equation}
         and the odd part is
         \begin{equation}\label{sec2_Lodd}
           L_{odd} = \sum_{n=0}^{n=\infty} L_{2n+1}.
         \end{equation}
         It will turn out that $n=0$ is sufficient to
         satisfy the BSFT condition (\ref{sec2_condition2relation}).

         We now introduce the parameter $\gamma$ to distinguish between
         the even ($\gamma=0$) and odd ($\gamma=1$) cases.
         Each term in the sums (\ref{sec2_Leven}) and (\ref{sec2_Lodd})
         goes as $T^{2n+\gamma}$:
         \begin{equation}\label{sec2_Lterms}
           L_{2n+\gamma} = \sum_{l=0}^{l=l_0}
             a_l^{(n)} (\partial_\mu T \partial^\mu T)^l T^{2n+\gamma-2l}
         \end{equation}

         It is known that string theory
         allows the ``S-brane'' solution \cite{SenRolling, SenMatter}
         \begin{equation}\label{sec2_Tsolution}
           T(t)=T_+ e^{t/\alpha}  + T_- e^{-t/\alpha}
         \end{equation}
         for the tachyon field $T$ in the spatially homogeneous case.
         We have here introduced another parameter, 
         $\beta \equiv \alpha^2$,
         which distinguishes between the bosonic ($\beta=1$) and the 
         superstring ($\beta=2$) case.

         The equations of motion which follow from 
         (\ref{sec2_L}), (\ref{sec2_Leven}) and (\ref{sec2_Lodd})
         are:
         \begin{equation}
           \begin{split}
             & \sum_{l=1}^{l=l_0} a_l^{(n)} 2l \partial^\mu 
               \left[
               (\partial_\lambda T \partial^\lambda T)^{l-1}
               (\partial_\mu T) T^{2n + \gamma -2l} 
               \right] \\
               =
             & \sum_{l=0}^{l=l_0} a_l^{(n)} (2n+\gamma-2l)
               (\partial_\mu T \partial^\mu T)^l T^{2n+\gamma-2l-1}
           \end{split}
         \end{equation}
         As indicated, these should be satisfied by the 
         solutions (\ref{sec2_Tsolution})
         for each $n$, as the solutions are exponentials.
         This renders the following recursion relation:
         \begin{equation}\label{sec2_recursionrel}
           a_{l+1}^{(n)} = \frac{\beta}{2}
           \frac{(2l-1)(2n+\gamma-2l)}{(2l+1)(l+1)} a_l^{(n)}
         \end{equation}
         We see from this that for the even case, $\gamma=0$,
         we get non-zero coefficients only for $l=0,\ldots,n$,
         i.e. the upper summation index  in (\ref{sec2_Lterms}) 
         is $l_0 = n$.
         For the odd terms, $\gamma=1$, we get
         non-zero coefficients for $l = 0,\ldots,\infty$, i.e.
         $l_0=\infty$.
         Note that the lower summation index is always 0, as
         already indicated in (\ref{sec2_Lterms}).

         The solution of the recursion relation is given by
         \begin{equation}\label{sec2_recursionsol}
           a_l^{(n)} = \left( \frac{\beta}{2} \right)^l
             \frac{1}{(2l-1)l!} 
             \frac{(2n+\gamma)!!}{(2n+\gamma-2l)!!}a_0^{(n).}
         \end{equation}
         In order to find the indicial coefficients $a_0^{(n)}$, it will
         be necessary to impose the boundary SFT (BSFT) condition, 
         to which we now turn.


         Using BSFT \cite{Witten1, Witten2, Sh1, Sh2},
         different tachyon profiles were used in
         \cite{KMM, KMM2, GS} to
         extract information about the tachyon Lagrangian, and
         rolling tachyons were studied within this framework e.g.
         in \cite{Minahan, ST}.

         According to BSFT
         \cite{Witten1, Witten2, Sh1, Sh2, Tseytlin1, Tseytlin2},
         the on-shell spacetime action $S_{on-shell}$ is,
         both in the bosonic and the superstring \cite{KMM2} case,
         proportional to the
         disc partition function $Z$:
         \begin{equation}\label{sec2_SFT}
           S_{on-shell} =c Z,
         \end{equation}
         where $c$ is a constant.
         The relation we will have use for is instead
         \begin{equation}\label{sec2_condition2relation}
           L_{on-shell} =c Z'(t),
         \end{equation}
         where $Z'(t)$ denotes the world-sheet disc partition function
         with the zero mode $t$ unintegrated.
         It is consistent in the sense that 
         (\ref{sec2_condition2relation}) 
         becomes (\ref{sec2_SFT}) upon
         integrating out the $t$ dependence.
         We will take $c=-1$, which is required for consistency
         with the results of
         \cite{LLM}
         (see also \cite{KutNi, Finn}).

         We will proceed by imposing this condition for the ``half S-brane''
         case
         $T_-=0$, i.e. $T=T_+ e^{t/\alpha}$.
         $Z'(t)$ is then given by \cite{Finn}
         \begin{equation}\label{sec2_Zprime}
           Z' = \frac{1}{1+\frac{1}{\beta} (2\pi)^\beta T^\beta}
              = \sum_{m=0}^{m=\infty} (-1)^m \frac{(2\pi)^{\beta m}}{\beta^m} T^{\beta m}.
         \end{equation}

         The terms (\ref{sec2_Lterms}) 
         in the on-shell Lagrangian, using (\ref{sec2_Tsolution}), are
         \begin{equation}
           L_{2n+\gamma}^{on-shell} = T^{2n+\gamma} \sum_{l=0}^{l=l_0} 
                          a_l^{(n)} \frac{(-1)^l}{\beta^l}.
         \end{equation}
         Imposing (\ref{sec2_condition2relation}) with $c=-1$, we see that
         relating equal powers of T requires
         \begin{equation}\label{sec2_m}
           m = \frac{1}{\beta} (2n+\gamma).
         \end{equation}
         Apparently, if $\beta=2$ (superstring case) then $\gamma=0$, i.e.
         only even terms are required. This means that $L_{super}$ will
         be invariant under $T \rightarrow -T$, as can be expected.
         If $\beta=1$ (bosonic case)
         no such restriction appears, i.e. we will need both even
         and odd terms.

         Matching terms in (\ref{sec2_condition2relation}) 
         corresponding to equal powers of $T$ gives the relation
         \begin{equation}\label{sec2_matching}
           \frac{(-1)^{m+1}}{\beta^m} = \sum_{l=0}^{l=l_0} a_l^{(n)} 
             \frac{(-1)^l}{\beta^l},
         \end{equation}
         where we have made the field redefinition
         $T \rightarrow 2\pi T$ to get
         rid of the factors of $2 \pi$.

         Since $a_l^{(n)}$ is proportional to $a_0^{(n)}$ according
         to (\ref{sec2_recursionsol}), this
         relation uniquely determines the indicial coefficient
         $a_0^{(n)}$, and that value can then be substituted back into
         (\ref{sec2_recursionsol}) to give the unique solution for
         the expansion coefficents $a_l^{(n)}$. 


         \subsection{Derivation}\label{sec2_derivation}


         Let us begin by treating the bosonic even case, which
         according to section \ref{sec2_conditions}
         corresponds to the parameter values
         $\beta=1$, $\gamma=0$, $l_0 = n$.
         Performing the sum in (\ref{sec2_matching}) using 
         (\ref{sec2_recursionsol}) and (\ref{sec2_m}) determines
         the indicial coefficent:
         \begin{equation}
           a_0^{(n)} = \frac{(2n-1)!!}{n! 2^n}
         \end{equation}
         Substituting this back into (\ref{sec2_recursionsol}) gives
         \begin{equation}
           a_l^{(n)} = \frac{(2n-1)!!}{2^n (2l-1) l! (n-l)!},
         \end{equation}
         which can be inserted into equations 
         (\ref{sec2_L}), (\ref{sec2_Leven})
         and (\ref{sec2_Lterms}) to give
         \begin{equation}\label{sec2_bos_even_Taylor}
           L_{bosonic}^{even} = \sum_{n=0}^{n=\infty} 
             \frac{(-1)^{n}(2n-1)!!}{2^n}
             \sum_{l=0}^{l=n} \frac{1}{(2l-1)(n-l)!l!} a^{n-l} b^l,
         \end{equation}
         where we have defined $a \equiv -T^2$ and
         $b \equiv -\partial_\mu T \partial^\mu T$.
         The lemma (\ref{app_lemma}) tells us that this is equal to
         \begin{equation}\label{sec2_bos_even}
           L_{bosonic}^{even} = -\frac{1}{1-T^2} \sqrt{1-T^2 - \partial_\mu T \partial^\mu T }.
         \end{equation}
         

         We now continue with the bosonic odd case, which corresponds to
         $\beta=1$, $\gamma=1$, $l_0 = \infty$.
         Performing the sum in (\ref{sec2_matching}) using 
         (\ref{sec2_recursionsol}) and (\ref{sec2_m}) determines
         the indicial coefficent:
         \begin{equation}
           a_0^{(n)} = -\frac{1}{\sqrt{\pi}} \frac{n!}{(n+\frac{1}{2})!}
         \end{equation}
         Substituting this back into (\ref{sec2_recursionsol}) gives
         \begin{equation}
           a_l^{(n)} = -\frac{1}{\sqrt{\pi}} \frac{n!}{(n+\frac{1}{2}-l)!}
             \frac{1}{(2l-1)l!}.
         \end{equation}
         Plugging this result into equations (\ref{sec2_L}), (\ref{sec2_Lodd})
         and (\ref{sec2_Lterms}) we get the bosonic odd Lagrangian,
         \begin{equation}\label{sec2_bos_odd_Taylor}
           L_{bosonic}^{odd} = -\frac{1}{\sqrt{\pi}} \sum_{n=0}^{n=\infty} n!
             \sum_{l=0}^{l=\infty} \frac{1}{(n+\frac{1}{2}-l)!(2l-1)l!}
             a^{2n+1} b^{l},
         \end{equation}
         where we have defined $a \equiv T$ and
         $b \equiv \frac{\partial_\mu T \partial^\mu T}{T^2}$.
         We contend that this equals (assuming $T=T(t)$)
         \begin{equation}\label{sec2_bos_odd}
           L_{bosonic}^{odd} = \frac{2}{\pi} \frac{1}{1-T^2}
           \left[
             \sqrt{1-T^2+\dot{T}^2} \sin^{-1} \left( 
               T\sqrt{1-\frac{\dot{T}^2}{T^2}}  \right)
             + \dot{T} \sin^{-1} \left( \frac{\dot{T}}{T} \right)
           \right].
         \end{equation}
         We believe that (\ref{sec2_bos_odd_Taylor}) and
         (\ref{sec2_bos_odd}) are identical for the following reason:
         According to section \ref{sec2_conditions}, 
         imposing the requirements that
         the equations of motion should be satisfied by
         (\ref{sec2_Tsolution}) and that (\ref{sec2_condition2relation}) 
         should hold uniquely determines the Lagrangian. Consequently, since
         we know that (\ref{sec2_bos_odd_Taylor}) fulfills these
         conditions, and it can be checked that (\ref{sec2_bos_odd})
         also satisfies these conditions, they must be identical.
         As a check, we have confirmed that the Taylor expansion of
         (\ref{sec2_bos_odd}) in $a$ and $b$ coincides
         with (\ref{sec2_bos_odd_Taylor}).


         To get the complete bosonic Lagrangian, we need to add
         the even (\ref{sec2_bos_even}) and odd 
         (\ref{sec2_bos_odd}) parts:
         \begin{equation}\label{sec2_Lbosonisk}
           \begin{split}
             L_{bosonic} & = L_{bosonic}^{even} + L_{bosonic}^{odd} \\
             & = -\frac{2}{\pi} \frac{1}{1-T^2} \left[
               \sqrt{1-T^2+\dot{T}^2} \cdot
               \cos^{-1} \left( T\sqrt{1-\frac{\dot{T}^2}{T^2}} \right)
               - \dot{T} \sin^{-1} \left( \frac{\dot{T}}{T}  \right)
             \right] \\
             & = +\frac{2}{\pi} \frac{1}{1-T^2}
             \left[
             \sqrt{T^2-\dot{T}^2-1} \cdot
             \cosh^{-1} \left( T \sqrt{1-\frac{\dot{T}^2}{T^2}} \right)
             + \dot{T} \sin^{-1} \left( \frac{\dot{T}}{T} \right)
             \right].
           \end{split}
         \end{equation}
         The first form is best suited for
         the field values $0 \le T^2-\dot{T}^2 \leq 1$, while
         the second is more appropriate when 
         $T^2-\dot{T}^2 \geq 1$ and positive $T$ (hence, $T \ge 1$).
         The forms are related by the relation
         $\cos^{-1} (1+\alpha^2) =  i \cosh^{-1}(1+\alpha^2)$.
         The forms for the bosonic Lagrangian given here assume
         that $T$ only depends on $t$, but since $\dot{T}$ always
         appears squared in terms of the Taylor expansion,
         covariance is manifest.


         Note that if we have a 
         half S-brane, i.e. if either $T_+$ or $T_-$ in
         (\ref{sec2_Tsolution}) vanishes, the Lagrangian
         (\ref{sec2_Lbosonisk})
         reduces on-shell to
         \begin{equation}\label{sec2_LhalfT}
           L_{bosonic}(\dot{T}=\pm T) =-\frac{1}{1+T}
         \end{equation}
         by virtue of (\ref{sec2_condition2relation}).
         For large $|T|$, $T$ and $\dot{T}$ are exponentially close
         modulo a sign,
         \begin{equation}\label{sec2_TTdot}
           \frac{\dot{T}}{T} = \pm 1 + O (e^{-2t}),
         \end{equation}
         using (\ref{sec2_Tsolution}). 
         Consequently,
         in the limit $|T| \rightarrow \infty$ the Lagrangian
         (\ref{sec2_Lbosonisk})
         behaves on-shell as
         \begin{equation}\label{sec2_LlargeT}
           L_{bosonic} \rightarrow -\frac{1}{T}.
         \end{equation}
         This coincides with the large $|T|$ behaviour of
         (\ref{sec2_LhalfT}),
         as expected since $|\dot{T}|=|T|$ holds exactly there.


         To get (\ref{sec2_Lbosonisk})
         we required that (\ref{sec2_Tsolution}) be a solution of
         the equations of motion. Let us mention that
         if we had instead imposed 
         $\tilde{T}=\sqrt{T_+} e^{t/2} + \sqrt{T_-} e^{-t/2}$ 
         as a solution for $\tilde{T}$ where $\tilde{T}^2 \equiv +T$
         (valid for $|T| \gg 2\sqrt{|T_+ T_-|}$),
         the resulting total bosonic Lagrangian in $T$, upon using the lemma
         (\ref{app_lemma}), would have been
         \begin{equation}\label{sec2_Ltilde}
           \tilde{L}_{bosonic} = 
             -\frac{1}{1+T} \sqrt{1 + T + 
             \frac{\partial_\mu T \partial^\mu T}{T} }.
         \end{equation}
         This Lagrangian, which contains both even and odd parts, 
         also has the properties (\ref{sec2_LhalfT}) and 
         (\ref{sec2_LlargeT}). Its equations of motion in $T$ are
         satisfied by the half S-brane but {\it not} by the full
         S-brane solution (\ref{sec2_Tsolution}), due to the large $T$
         approximation.
         Making the field redefinition
         \begin{equation}\label{sec2_field_redef_bosonic}
           T \equiv \sinh^2 \left( \frac{\tilde{T}}{2} \right)
         \end{equation}
         takes (\ref{sec2_Ltilde}) to the even standard form 
         (\ref{sec1_standard})
         for $T \ge 0$.

         
         Finally, let us also mention that the superstring case
         (non-BPS brane in type II string theory),
         corresponding to $\beta=2$, $\gamma=0$, $l_0 = n$,
         can also
         be treated in an analogous way, and again using the lemma
         (\ref{app_lemma}), the result is \cite{KutNi}
         \begin{equation}\label{sec2_Lsuper}
           L_{super} = -\frac{1}{1+\frac{1}{2}T^2} \sqrt{1+\frac{1}{2}T^2
             +\partial_\mu T \partial^\mu T }.
         \end{equation}
         The field redefinition
         \begin{equation}
           T \equiv \sqrt{2} \sinh \left( \frac{\tilde{T}}{\sqrt{2}} \right)
         \end{equation}
         then transforms the Lagrangian (\ref{sec2_Lsuper}) into the standard form
         (\ref{sec1_standard}).

         \section{Properties of the Bosonic Lagrangian}\label{sec3}
         \setcounter{equation}{0} 

         In the superstring case, the Lagrangian (\ref{sec1_standard})
         has proven to satisfy several different consistency checks,
         so in this section we will investigate some of the properties
         of (\ref{sec2_Lbosonisk});
         the energy-momentum tensor in section \ref{sec3_EMtensor},
         and the potential, closed string vacuum and lumps in
         section \ref{sec3_other}.


        \subsection{Energy-Momentum Tensor}\label{sec3_EMtensor}


         We will now compare the behaviour of the
         energy-momentum tensor of the theory defined by
         (\ref{sec2_Lbosonisk}) to that of the BCFT analysis
         \cite{SenRolling, SenMatter}.


         We begin by discussing the full S-brane case $\rho < \tau_p$
         for which 
         the BCFT energy-momentum tensor is given by
         \cite{SenRolling, SenMatter}       
         \begin{equation}\label{sec3_SenT1}
           \begin{split}
             \rho_{BCFT} & = \tau_p \cos^2(\pi \tilde{\lambda}) \\
             (p_{BCFT})_i(t) & = -\tau_p \left[ \frac{1}{1+\sin(\pi \tilde{\lambda}) e^t}
                                 + \frac{1}{1+\sin(\pi \tilde{\lambda}) e^{-t}}
                                 -1 \right]
           \end{split},
         \end{equation}
         where $i=1,\ldots,p$.
         The constant $\tilde{\lambda}$ parameterizes the solutions,
         and should satisfy\footnote{
         Recall that we rescaled $T$ in equation (\ref{sec2_matching}).
         It is really $\hat{T}=\frac{1}{2\pi}T$ which satisfies
         the BSFT condition, hence $\hat{T_0}=\frac{1}{2\pi}T_0$.}
         $\tilde{\lambda}=\hat{T}_0+O(\hat{T}_0^2)$, i.e.
         \begin{equation}\label{sec3_parameter}
           \tilde{\lambda} = \frac{1}{2\pi} T_0 + O(T_0^2),
         \end{equation}
         where $T_0$ is the turning point of the tachyon,
         where it has potential energy only:
         (\ref{sec2_Tsolution}) becomes
         \begin{equation}\label{sec3_Tsol}
           T=T_0 \cosh(t),
         \end{equation}
         which is related by a time translation to the solution 
         (\ref{sec2_Tsolution}) with $T_0^2 = + 4T_+ T_-$,
         because for this case, $T^2 - \dot{T}^2 = 4T_+ T_- > 0$.

         The energy-momentum tensor is defined and calculated as
         \begin{equation}\label{sec3_EMtensor_def}
           T_{\mu \nu} \equiv -\frac{2}{\sqrt{-\eta}} \frac{\delta S}{\delta \eta^{\mu \nu}},
         \end{equation}
         with our Lagrangian given by
         (\ref{sec2_Lbosonisk}).
         The on-shell energy (\ref{sec3_EMtensor_def}) is then
         \begin{equation}\label{sec3_energy}
           \rho \equiv T_{tt} = V(T_0),
         \end{equation}
         with the potential $V(T_0)$ given in (\ref{sec3_potential}).
         In (\ref{sec3_SenT1}), the energy
         is invariant under $\tilde{\lambda} \rightarrow -\tilde{\lambda}$,
         but (\ref{sec3_energy}) is odd for small $T_0$;
         physical tachyon energies have the wrong sign for $T_0<0$.
         For $T_0 \ge 0$, (\ref{sec3_energy})
         satisfies
         $0 \le \rho \le 1 = \tau_p$,
         (using (\ref{sec3_energydensity})), as expected from 
         (\ref{sec3_SenT1}), and
         at $T_0 = \tilde{\lambda}= 0$ we get
         $\rho=\rho_{BCFT}=+1$.

         Turning to the pressure, we get from (\ref{sec3_EMtensor_def}) that
         \begin{equation}\label{sec3_pressure}
           p_i(t) \equiv T_{ii} =  L_{bosonic},
         \end{equation}
         where $i=1,\ldots,p$ and
         $L_{bosonic}$ is given in (\ref{sec2_Lbosonisk}).
         The numerator $N$ of $p_i(t)=L_{bosonic}$ in (\ref{sec2_Lbosonisk})
         on-shell is
         \begin{equation}\label{sec3_N}
           N = f(T_0) - T f\left( \frac{T_0}{T} \right).
         \end{equation}
         The function $f$ is defined by
         \begin{equation}\label{sec3_f}
           f(x) \equiv \sqrt{1-x^2} \cos^{-1}(x)
                \equiv \sum_{n=0}^{n=\infty} a_n x^n,
         \end{equation}
         where we also defined the Taylor expansion coefficients $a_n$
         for $f(x)$.
         The numerator $N$ vanishes for $T=1$,
         but not for $T=-1$, so the numerator may contain a factor
         of $(T-1)$. More explicitly, inserting
         the indicated Taylor expansion of $f(x)$ gives
         \begin{equation}\label{sec3_NinG}
           N = (T-1) \sum_{n=0}^{n=\infty} a_n \frac{T_0^n}{T^{n-1}} G_{n-1}(T),
         \end{equation}
         where $G_n(x) \equiv \frac{x^n-1}{x-1} = 1+x+x^2 + \cdots + x^{n-1} $ 
         denotes the
         geometric sum with $n$ terms.

         Consider now the well-behaved side $T \ge 0$.
         The pressure $p_i(t)$ is
         regular at $T=+1$, since the numerator contains a factor of
         $(T-1)$ (see (\ref{sec3_NinG})).
         As the tachyon rolls towards the closed string vacuum
         $T \rightarrow \infty$
         we get, using
         (\ref{sec2_LlargeT}) in (\ref{sec3_pressure}), that 
         \begin{equation}\label{sec3_pressure_largeT}
           p_i(t) = -\frac{1}{T}
             \longrightarrow 0.
         \end{equation}
         Thus, 
         the pressure vanishes as it should
         according to (\ref{sec3_SenT1}), using (\ref{sec3_energydensity})
         and making the
         identification
         $\pi \tilde{\lambda} = \frac{T_0}{2} + O(T_0^2)$
         for sufficiently small $T_0$, which matches the expectation
         (\ref{sec3_parameter}).

         Consider next the singular side $T \le 0$.
         As expected from (\ref{sec3_SenT1}),
         as the tachyon rolls down towards the
         singularity at $T=-1$, the pressure blows up
         in a finite time:
         The discussion after equation (\ref{sec3_f}) tells us
         that the numerator (\ref{sec3_N}) of $L_{bosonic}$
         in (\ref{sec2_Lbosonisk})
         does not vanish as $T \rightarrow -1$,
         so $p_i(t) = L_{bosonic}$ has a simple pole there.

         Now, consider what happens around $T=0$.
         For small $T_0$, equation (\ref{sec3_N}) becomes
         \begin{equation}
           N = (1-T) \left(\frac{\pi}{2}-T_0 \right) + O(T_0^2).
         \end{equation}
         Hence, the pressure is
         \begin{equation}\label{sec3_pressure0}
           \begin{split}
             p_i(t) & = -\left[ 1-\frac{2}{\pi}T_0 \right]
                       \frac{1}{1+T} +O(T_0^2) \\
                    & = -\tau_p \left( 1-\frac{2}{\pi}T_0 \right)
                       \left[ \frac{1}{1+\frac{T_0}{2} e^t}+\frac{1}{1+\frac{T_0}{2} e^{-t}}-1 \right]
                       + O(T_0^2).
           \end{split}
         \end{equation}
         using $\tau_p=+1$ from (\ref{sec3_energydensity}).
         We see that the $t$ dependence remains the same as in
         (\ref{sec3_SenT1}) to order $O(T_0^2)$, provided that
         we replace $\tau_p$
         by $\tau_p \left( 1 - \frac{2}{\pi}T_0 \right)$,
         and make the identification
         $\pi \tilde{\lambda} = \frac{T_0}{2} + O(T_0^2)$,
         again in agreement with (\ref{sec3_parameter}).

         We will make one final comment before leaving the case
         $\rho < \tau_p$:
         Placing the bosonic tachyon at rest at the closed string vacuum,
         $\tilde{\lambda}=\frac{1}{2}$,
         gives vanishing pressure and energy 
         (\ref{sec3_SenT1}),
         and the same is true for (\ref{sec3_energy}) and 
         (\ref{sec3_pressure_largeT}), because the initial conditions
         $T=\infty$, $\dot{T}=0$ in (\ref{sec2_Tsolution})
         corresponds to $T_+ = T_- = \infty$, i.e. $T_0 = \infty$.


         We now turn to the case $\rho > \tau_p$.
         For $|\sinh(\pi \tilde{\lambda})|<1$, 
         the BCFT energy-momentum tensor is given by
         \cite{SenRolling, SenMatter}       
         \begin{equation}\label{sec3_SenT2}
           \begin{split}
             \rho_{BCFT} & = \tau_p \cosh^2(\pi \tilde{\lambda}) \\
             (p_{BCFT})_i(t) & = -\tau_p \left[ \frac{1}{1+\sinh(\pi \tilde{\lambda}) e^t}
                                 + \frac{1}{1-\sinh(\pi \tilde{\lambda}) e^{-t}}
                                 -1 \right]
           \end{split},
         \end{equation}
         where $i=1,\ldots,p$ and
         $\tilde{\lambda}$ parameterizes the solutions.
         Now, $T^2 -\dot{T}^2 = 4T_+ T_- < 0$ on-shell using
         the solution (\ref{sec2_Tsolution}),
         which can be time translated to
         \begin{equation}\label{sec3_Tsol2}
           T=T_0 \sinh(t),
         \end{equation}
         where $T_0^2 = - 4T_+ T_-$;
         the tachyon starts at $T=0$
         with a velocity $\partial_t T = T_0$.
         As will be discussed in section \ref{sec3_other},
         the kinetic term blows up at $T=0$, so the
         energy-momentum tensor cannot be well-defined;
         in general, it has imaginary parts (cf. section \ref{sec4}).
         Nevertheless, a similar analysis as the one
         for the case $\rho < \tau_p$ results in
         a conserved energy and a pressure which blows up at
         $T=-1$ and decreases as it should (\ref{sec3_SenT2})
         for large $T$.


         Let us now complete the discussion of the
         energy-momentum tensor by analysing the half S-brane
         $\rho = \tau_p$, corresponding to
         starting the tachyon off at rest displaced slightly
         from the top of the potential hill.
         The BCFT energy-momentum
         tensor for this case is\footnote{
         For
         concreteness, we study the case
         $T \sim e^t$, but the case $T \sim e^{-t}$
         is handled similarly.}
         \cite{Str3, Str4, LLM, Finn}
         \begin{equation}\label{sec3_SenT3}
           \begin{split}
             \rho & = \tau_p \\
             p_i(t) & = -\tau_p \frac{1}{1+\hat{\lambda}e^t}
           \end{split},
         \end{equation}
         where $\hat{\lambda}$
         is not really a parameter in this case;
         it can be set to unity by time translation.
         The energy-momentum tensor of our theory is precisely equal
         to the BCFT one, using $\tau_p = 1$ from (\ref{sec3_energydensity})
         and making the identification $\hat{\lambda} = T_+$:
         The energy can be checked by setting $T_0 = 0$
         in (\ref{sec3_energy}).
         The pressure is also correct, as can be seen by
         combining (\ref{sec3_pressure}) with (\ref{sec2_LhalfT}).
         This is not surprising since (\ref{sec2_LhalfT}) is a
         direct consequence of
         having imposed the condition
         (\ref{sec2_condition2relation}).
         The BCFT properties that the pressure is singular at $T=-1$
         and falls off exponentially
         for $T \rightarrow \infty$ therefore carry over automatically
         to our theory in this case.

         \subsection{Other properties}\label{sec3_other}

\begin{figure}[htb]
           \centering
           \includegraphics[height=6cm,width=6cm]{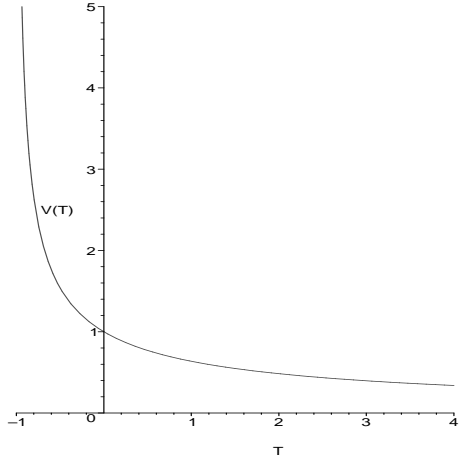}
           \caption{Potential of equation (\ref{sec3_potential})}
           \label{figur_potential}
         \end{figure}

         Let us begin by examining the potential, which follows immediately 
         from (\ref{sec2_Lbosonisk}):
         \begin{equation}\label{sec3_potential}
           V(T) = \frac{2}{\pi} \frac{1}{\sqrt{1-T^2}} \cos^{-1}(T)
                = \frac{2}{\pi}\frac{1}{\sqrt{T^2-1}}\cosh^{-1}(T),
         \end{equation}
         where the first form is best for $|T|<1$, and the second
         is better for $T>1$.
         The potential is shown in figure \ref{figur_potential}.
         Note that at $T=-1$ the potential explodes to $+\infty$,
         and that
         $V(T\rightarrow \infty) \longrightarrow 0$.
         The energy density of the D-brane at the open-string vacuum
         can be defined and computed as
         \begin{equation}\label{sec3_energydensity}
           \tau_p \equiv V(0) = 1.
         \end{equation}
         The odd contribution makes the potential monotonically decreasing.
         In particular, at $T=0$, it is proportional to $-T + O(T^2)$
         (modulo a constant vacuum energy due to the D-brane)
         instead of the more traditional $-T^2 + O(T^2)$ for a maximum.
         To still allow for the solutions (\ref{sec2_Tsolution}),
         the kinetic term blows up\footnote{
         In a sense, the fact that the kinetic term 
         blows up at $T=0$
         is the ``glue'' that makes a tachyon placed at rest at $T=0$
         stay there.} 
         and changes sign at $T=0$,
         making the initial-value problem for rolling tachyons with
         $\rho > \tau_p$ not well-defined. There cannot therefore be
         a field redefinition which takes (\ref{sec2_Lbosonisk})
         to the canonical form even locally; if there were, 
         (\ref{sec2_Tsolution}) would not have been restricted to
         $T_+ T_- \ge 0$. Still, the solution 
         (\ref{sec2_Tsolution}) tells us that
         (\ref{sec2_Lbosonisk}) describes a tachyon of mass $m^2=-1$.

         In view of the potential (\ref{sec3_potential}) and the
         properties of the energy-momentum tensor
         discussed in section \ref{sec3_EMtensor}, it appears
         that the following regions can be identified:
         \begin{itemize}
           \item $T=-1$ is the singularity where the 
                        pressure blows up.
           \item $T=0$ is the open-string vacuum (since (\ref{sec2_Tsolution})
                       has been imposed as a solution).
           \item $T=+\infty$ is the closed-string vacuum.
         \end{itemize}


         An important
         expected property of the closed-string vacuum is that
         it should not admit any open-string particle excitations.
         As discussed in \cite{SenField}, there are basically two
         ways in which this property can be realized. In some field
         theory models
         \cite{FTM1, FTM2, FTM3}
         the second derivative of the tachyon potential blows up
         at the minimum, hence rendering infinite mass to any
         would-be plane-wave excitations.
         The second possibility, which we will investigate here analogously
         to what was done in \cite{SenField},
         is that the absence of plane-wave
         solutions is due to the behaviour of the kinetic term
         around the minimum (i.e. $T=\infty$), 
         as in \cite{FTM4, FTM5, FTM6, FTM7}.

         First, note that in the large $T$ limit, 
         the Lagrangian
         (\ref{sec2_Lbosonisk}) becomes          
         \begin{equation}\label{sec3_Lapprox}
           -\frac{\pi}{2} L = \frac{1}{T} \left[ \ln T\left(
           1+\frac{1}{2} \frac{\partial_{\mu}T \partial^{\mu}T -1}{T^2} \right) - \frac{\partial_{\mu}T\partial^{\mu}T}{T^2} \right] + \cdots.
         \end{equation}         
         Making the field redefinitions
         \begin{equation}\label{sec3_fieldredef}
           \begin{split}
                T & \equiv (1+\tilde{T}) e^{\tilde{T}} \\
             \phi & \equiv e^{-\tilde{T}/2}
           \end{split}
         \end{equation}
         turns (\ref{sec3_Lapprox}) into
         \begin{equation}\label{sec3_Linphi}
           L = \frac{8}{\pi} \left( -\frac{1}{2} \partial_{\mu}\phi \partial^{\mu}\phi - \frac{1}{4} \phi^2  
               \right) + \cdots.
         \end{equation}
         In terms of $\phi$, which expands around $T=\infty$, 
         the Lagrangian has the right canonical
         form, so this is the field definition to use for analysing
         excitations around the closed string vacuum.
         The Hamiltonian
         \begin{equation}
           H = \frac{2}{\pi} 
               \frac{1}{\sqrt{T^2+\partial_{\mu}T \partial^{\mu}T-1}} cosh^{-1} \left[
               T\sqrt{1+\frac{\partial_{\mu}T \partial^{\mu}T}{T^2}} \right],
         \end{equation}
         corresponding to the Lagrangian (\ref{sec2_Lbosonisk}),
         should be constant for solutions of the equations of motion,
         which means that
         \begin{equation}\label{sec3_COM}
           T^2 + \partial_{\mu} T \partial^{\mu}T = C
         \end{equation}
         for some constant $C$.
         However, the plane-wave ansatz
         \begin{equation}
           \phi = ae^{ik \cdot x}
         \end{equation}
         makes the left-hand side of (\ref{sec3_COM}) become, 
         using (\ref{sec3_fieldredef}),
         \begin{equation}
           \frac{1}{\phi^4} \left[ (1-2 \ln \phi)^2 - 16k^2 (1-\ln \phi)^2 \right],
         \end{equation}
         which is not constant for any $k^2 = -m^2$.
         We are assuming a flat closed string background, so the
         concept of particle excitations in itself is well-defined.
         We therefore conclude that open-string particle excitations
         are absent at the closed-string vacuum $T=\infty$,
         as promised. 
         

         Finally, let us discuss the viability
         of lump solutions.
         In the superstring case, the Dirac-Born-Infeld action
         has been shown to allow kink solutions \cite{SenKink}
         (stable D-brane configurations).
         In the bosonic case, there should instead
         exist lump solutions (unstable D-branes). 
         However, the Lagrangian (\ref{sec2_Lbosonisk})
         is designed to be valid for profiles in the vicinity of 
         rolling tachyons (cf. section \ref{sec4}),
         and we can actually show that it does not
         contain lumps of the type locally resembling the kink solutions
         of \cite{SenKink}.

         By analogy with the superstring case \cite{SenKink}, 
         let us begin by analysing the 
         energy-momentum tensor, which in the bosonic time-independent
         case $T=T(x)$ becomes
         \begin{equation}\label{sec3_Tinx}
         \begin{split}
           T_{\alpha \beta} & = \eta_{\alpha \beta} L_{bosonic} \\
           \rho \equiv T_{tt} & = - L_{bosonic} \\
           p \equiv T_{xx} & =  -\frac{2}{\pi}\frac{1}{\sqrt{1-T^2-{T'}^2}}\cos^{-1}
             \left( T \sqrt{ 1 + \frac{{T'}^2}{T^2}} \right) \\
           & = -\frac{2}{\pi} \frac{1}{\sqrt{T^2+{T'}^2-1}}
             \cosh^{-1} \left( T\sqrt{1+\frac{{T'}^2}{T^2}} \right)
         \end{split},
         \end{equation}
         where $\alpha$,$\beta=0,\ldots,p-1$, $x^p \equiv x$.
         The pressure $p$ is given in two forms, depending
         on whether $T^2 + {T'}^2 \le 1$ or $T^2 + {T'}^2 \ge 1$, 
         respectively.         
         Since we are looking for time-independent lump solutions,
         $T=T(x)$, we should substitute
         $\dot{T} \rightarrow \pm iT'$ in the Lagrangian
         (\ref{sec2_Lbosonisk}).
         The conservation law $\partial^\alpha T_{\alpha \beta} =0$
         now reduces to
         \begin{equation}
           \frac{\partial p}{\partial x} = 0.
         \end{equation}
         Using (\ref{sec3_Tinx}), this means that to have a solution we
         need to have either\footnote{
         This solution may also correspond
         to D-branes in the world-sheet theory \cite{HKM}, but we
         do not analyse it further here.} $T = A \cos(x) + B \sin(x)$
         (which is the Lorentz rotated version of (\ref{sec2_Tsolution})),
         or
         \begin{equation}\label{sec3_lump_equation}
           T^2 + {T'}^2 \rightarrow \infty.
         \end{equation}         
         As in \cite{SenKink}, it is clear that for example a linear
         profile of infinite slope would locally satsify 
         (\ref{sec3_lump_equation}) for $T>-1$. 
         However, we can never find a profile
         that satisfies (\ref{sec3_lump_equation}) globally, since
         a lump coming in from $T=+\infty$ must eventually turn at 
         some finite $T$.
         Analysing lumps therefore seems to require higher derivative
         corrections to the Lagrangian (\ref{sec2_Lbosonisk}).


         \section{Discussion}\label{sec4}
         \setcounter{equation}{0} 
      
         Our main result is the effective Lagrangian
         (\ref{sec2_Lbosonisk})
         for the bosonic open-string tachyon. It is the unique
         covariant Lagrangian not containing second or higher
         order derivatives whose equations of motion are satisfied
         by the solutions (\ref{sec2_Tsolution})
         and which on-shell is proportional to the disc partition
         function (with zero modes unintegrated) on half S-branes.

         In the superstring case (non-BPS brane in type II string theory), 
         this procedure leads to 
         (\ref{sec2_Lsuper})
         which is related by a field redefinition to the standard form
         (\ref{sec1_standard}). According to \cite{KutNi}, it describes 
         tachyon profiles of the type
         \begin{equation}\label{sec4_validityregion}
           T=T_+ (x^\mu) e^{t/\alpha} + T_- (x^{\mu}) e^{-t/\alpha}
         \end{equation}
         (with $\alpha=\sqrt{2}$) 
         where $|\partial_\mu T_+| \ll 1$ and $|\partial_\mu T_-| \ll 1$,
         expanded around a solution with $T_-=0$ (or $T_+=0$).
         In particular, it has one free parameter $a_0^{(n)}$ 
         at each order $T^{2n}$ (see (\ref{sec2_recursionsol})),
         precisely as the $2n$-point function of tachyons
         at order momentum squared in a spatially homogeneous background
         of half S-branes (i.e. $T_-=0$).
         The bosonic Lagrangian (\ref{sec2_Lbosonisk})
         should then analogously be designed to describe the same type of
         profiles (\ref{sec4_validityregion}) 
         (with $\alpha=1$) for all values of $T>-1$, 
         provided that $|\dot{T}| \le |T|$. 
         We will explain these conditions shortly.

         We also noted in section \ref{sec2_derivation} 
         that the bosonic Lagrangian 
         (\ref{sec2_Ltilde}) could be derived by a similar procedure
         valid for $|T| \gg 2\sqrt{|T_+ T_-|}$ 
         (see the discussion around equation 
         (\ref{sec2_Ltilde})), and it therefore describes profiles
         of the type (\ref{sec4_validityregion}) when
         $|T| \gg 2\sqrt{|T_+(x^\mu) T_-(x^\mu)|}$.
         Since it was related to the standard form (\ref{sec1_standard})
         by the field redefinition (\ref{sec2_field_redef_bosonic}),
         it appears that (\ref{sec1_standard}) can only be expected
         to describe bosonic profiles well in the vicinity of rolling tachyons
         for large positive $T$
         (cf. the discussion in \cite{SenTT}).

         The most important test of (\ref{sec2_Lbosonisk}) 
         was the behaviour of the physical
         energy-momentum tensor (section \ref{sec3_EMtensor}).
         The pressure blew up in a finite time for tachyons rolling
         towards the singular side of the potential, and at the other
         side the pressure eventually decreased exponentially, as
         expected.
         Quantitative agreement was also good for half S-branes
         or full S-branes with either $T_0 \rightarrow 0$ or
         $T \rightarrow \infty$, i.e. close to half S-branes, which
         of course lies in line with the expectation 
         (\ref{sec4_validityregion}).

         In general, the Lagrangian (\ref{sec2_Lbosonisk}), and therefore also
         the physical energy-momentum tensor (cf. section 
         \ref{sec3_EMtensor}), acquires imaginary parts
         outside of the region
         $T \ge -1$ and $|\dot{T}| \le |T|$.
         The former restriction corresponds to going past the singularity
         at $T=-1$
         where the pressure blows up into the unphysical region $T<-1$. 
         The latter restriction
         is related to the fact that for rolling tachyons of the 
         type $T = T_0 \sinh(t)$,
         the kinetic term
         blows up at $T=0$, i.e. tachyons which traverse the top of the 
         potential
         hill cannot be assigned a well-defined energy. 
         The singularity of the kinetic term
         can be traced back to the fact that the EOM condition in
         section \ref{sec2_conditions} made the Lagrangian
         non-analytic in $T$.
         Indeed, the Taylor expansion in
         $b=-\left(\frac{\dot{T}}{T}\right)^2$,
         given in (\ref{sec2_bos_odd_Taylor}),
         diverges by the quotient test for $|\dot{T}| > |T|$,
         leading us to restrict the validity region to
         $|\dot{T}| \le |T|$.
         Analytic continuation beyond this region, while possible,
         introduces imaginary components\footnote{
         However, note that static profiles $T=T(x)$ are not restricted
         to $|T'| \le |T|$.}.

         Imposing the BSFT condition 
         (\ref{sec2_condition2relation}) for full S-branes
         would rid the Lagrangian of imaginary parts.
         Since the freedom of the ansatz
         $L(T, \partial_\mu T \partial^\mu T)$ was saturated already
         by half S-branes, this would require adjusting the ansatz
         such that more freedom is provided. The right type of
         freedom could possibly be provided by allowing
         for higher derivatives. This would also
         change the equations of motion, possibly in such a way
         so as to remove the non-analyticity in $T$. 

         However, 
         perhaps there is another
         ansatz than $L=L(T,\partial_\mu T \partial^\mu T)$ which contains
         the same amount of freedom.
         Suppose, for example, that we make the ansatz 
         $L=L(T,\partial_\mu T \partial^\mu T, \partial^2 T)$.
         Instead of (\ref{sec2_Lterms}), we then get
         \begin{equation}\label{sec4_Lterms}
           L_{2n+\gamma} = \sum_{l=l_0}^{l=l_1} \sum_{m=m_0}^{m=m_1}
             a_{lm}^{(n)} (\partial_\mu T \partial^\mu T)^l
             (\partial^2 T)^m
             T^{2n+\gamma-2l-m}
         \end{equation}
         at order $T^{2n+\gamma}$.
         Requiring that the equations of motion are consistent as
         described in section \ref{sec2_conditions} leads to
         \begin{equation}\label{sec4_recursionrel}
           \begin{split}
          0 & = a_{l+1,m}^{(n)} 2(m-1)(l+1)(2l+1) \\
            & + a_{l,m}^{(n)} \left[ (2n+\gamma-2l)2l(m-1)+
                 m(2n+\gamma-2l-1)(2l+1) + (2n+\gamma-2l-m)  \right] \\
            & + a_{l-1,m}^{(n)} m (2n+\gamma-2l+1)(2n+\gamma-2l)
           \end{split}.
         \end{equation}
         The case $m=0$ is the case already treated in section 
         \ref{sec2_conditions}, 
         i.e. $a_{l,0}^{(n)}=a_l^{(n)}$ in (\ref{sec2_recursionrel})
         (with only one free parameter, $a_0^{(n)}$).
         But we see from (\ref{sec4_recursionrel}) that there is another
         possibility, $m=1$, which also has only one free parameter at
         each order. This example therefore shows that there
         may be higher derivative Lagrangians which also describe
         profiles of the type (\ref{sec4_validityregion}), possibly
         without the restriction $|\dot{T}| \le |T|$.


         \bigskip
         \noindent
        {\bf Acknowledgments:} I am very grateful to J.\ Minahan
        for several enlightening discussions and many useful comments on the
        manuscript. I would also like to thank
        L.\ Freyhult, J.\ Gregory, F.\ Kristiansson, U.\ Lindstr\"om
        and V.\ Schomerus for conversations.

         \begin{appendix} 

         \section{Appendix}
         \setcounter{equation}{0} 

         We want to show that
         \begin{equation}\label{app_lemma}
             L \equiv \frac{1}{1+a} \sqrt{1+a+b}
             = \sum_{n=0}^{n=\infty} \frac{(-1)^{n+1}(2n-1)!!}{2^n}
               \sum_{l=0}^{l=n}\frac{1}{l!(n-l)!(2l-1)}a^{n-l}b^l.
         \end{equation}
         This formula was basically derived as a part of \cite{KutNi},
         but since it is used several times in section
         \ref{sec2_derivation}, we
         repeat the derivation here.
         Begin by Taylor expanding the left-hand side in $a$ and $b$ and
         use that $(-1)! = (-2)! = \cdots = \infty$ to drop terms give
         \begin{equation}
           L = \sum_{n=0}^{n=\infty} (-1)^n
           \sum_{m=0}^{\infty} \frac{(2m-3)!!(-1)^{m+1}}{2^m}
           \sum_{k=0}^{k=p} \frac{1}{k!(m-k)!} a^{n+m-k}b^k.
         \end{equation}
         Upon substituting $p \equiv n+m$,
         \begin{equation}
           L =
           \sum_{p=0}^{p=\infty} (-1)^{p+1} \sum_{k=0}^{k=p} \frac{1}{k!}
           \sum_{m=0}^{m=p} \frac{(2m-3)!!}{2^m (m-k)!} a^{p-k} b^k.
         \end{equation}
         The last term can be rewritten using $m \rightarrow m+k$
         and the identity
         \begin{equation}
           \sum_{m=0}^{m=p-k} \frac{\left[ 2(m+k) -3 \right]!!}{2^m m!}
           = \frac{2^{k-p}(2p-1)!!}{(p-k)!(2k-1)},
         \end{equation}
         which collapses one of the sums.
         This leaves us with
         \begin{equation}
           L = \sum_{p=0}^{p=\infty} \frac{(-1)^{p+1}(2p-1)!!}{2^p}
             \sum_{k=0}^{k=p} \frac{1}{k!(p-k)!(2k-1)} a^{p-k}b^k,
         \end{equation}
         which is the same as the right-hand side of (\ref{app_lemma}), upon 
         replacing $p \rightarrow n$
         and $k \rightarrow l$.

         \end{appendix}

         \end{document}